\documentclass[a4paper,11pt]{article}
\usepackage{pos}
\usepackage{amsmath}
\usepackage{bbold}
\usepackage{adjustbox}
\usepackage{float}
\usepackage{multirow}
\usepackage{comment}

\title{$\bar{b}\bar{b}ud$ Tetraquarks with $I(J^P)=0(1^-)$ and $\bar{b}\bar{c}ud$ Tetraquarks with $I(J^P)=0(0^+)$ and $I(J^P)=0(1^+)$ from Lattice QCD Antistatic-Antistatic Potentials}
\ShortTitle{$\bar{b}\bar{b}ud$ and $\bar{b}\bar{c}ud$ Tetraquarks from Lattice QCD Antistatic-Antistatic Potentials}

\newcommand{\tens}[1]{%
  \mathbin{\mathop{\otimes}\limits_{#1}}%
}

\author*[a]{Jakob Hoffmann}
\author[a] {Lasse Müller}
\author[a,b]{Marc Wagner}

\affiliation[a]{Goethe-Universität Frankfurt am Main, Institut für Theoretische Physik, Max-von-Laue-Straße 1, D-60438 Frankfurt am Main, Germany}

\affiliation[b]{Helmholtz Research Academy Hesse for FAIR, Campus Riedberg, Max-von-Laue-Straße 12, \\ D-60438 Frankfurt am Main, Germany}

\emailAdd{jhoffmann@itp.uni-frankfurt.de}
\emailAdd{lmueller@itp.uni-frankfurt.de}
\emailAdd{mwagner@itp.uni-frankfurt.de}

\abstract{We study heavy spin effects in $\bar{b}\bar{b}ud$ and $\bar{b}\bar{c}ud$ four-quark systems using the Born-Oppenheimer approximation and existing antistatic-antistatic potentials computed with lattice QCD. We report about a recent refined investigation of the $\bar{b}\bar{b}ud$ system with $I(J^P)=0(1^-)$, where we predicted a tetraquark resonance slightly above the $B^{*}B^{*}$ threshold. Furthermore, we extend our Born-Oppenheimer approach to $\bar{b}\bar{c}ud$ four-quark systems. For quantum numbers $I(J^P)=0(0^+)$ as well as $I(J^P)=0(1^+)$ we find virtual bound states rather far away from the lowest meson-meson thresholds.}

\FullConference{The 41st International Symposium on Lattice Field Theory (LATTICE2024)\\
 28 July - 3 August 2024\\
Liverpool, UK}

\begin{document}

\maketitle


\section{Introduction}

In this talk we discuss our investigations of $\bar{b}\bar{b}ud$ and $\bar{b}\bar{c}ud$ four-quark systems using the Born-Oppenheimer approximation, which is a two-step approach. In the first step antistatic-antistatic potentials in the presence of two light quarks are computed with lattice QCD (see Section~\ref{SEC_V5_Vj}). In the second step, these potentials are used in appropriate coupled-channel Schr\"odinger equations, where bound states as well as resonances can be predicted using standard techniques from non-relatvistic quantum mechanics (see Section~\ref{SEC_SG1} and Section~\ref{SEC_SG2}). Using such coupled-channel Schr\"odinger equations as well as experimental results for $B$, $B^\ast$, $D$ and $D^\ast$ mesons allows to take into account effects from the heavy quark spins, even though the antistatic-antistatic potentials are degenerate with respect to these spins.

In the following sections we briefly summarize a completed study of a $\bar{b}\bar{b}ud$ tetraquark resonance with quantum numbers $I(J^P) = 0(1^-)$, where details have recently been published in Ref.\ \cite{Hoffmann:2024hbz}. We also present theoretical basics and new results for $\bar{b}\bar{c}ud$ four-quark systems with quantum numbers $I(J^P)=0(0^+)$ as well as $I(J^P)=0(1^+)$, which have previously not been investigated within the Born-Oppenheimer approximation.


\section{\label{SEC_V5_Vj}The Antistatic-Antistatic-Light-Light Potentials $V_5(r)$ and $V_j(r)$}

\begin{figure}[htb]
    \centering
    \includegraphics[width=0.5\linewidth]{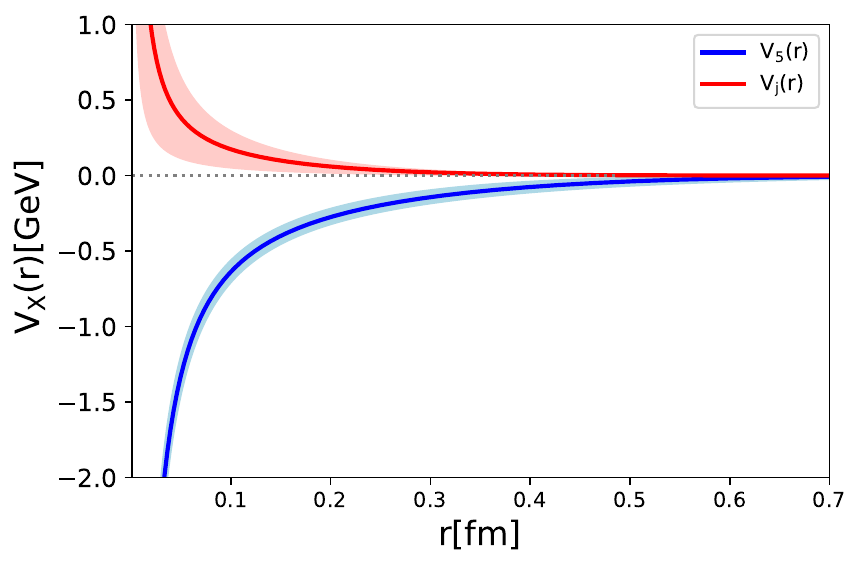}
    \caption{Parametrizations of lattice QCD results from Ref~[2]
    for the $\bar{Q} \bar{Q} q q$ potentials $V_5(r)$ and $V_j(r)$.}
    \label{fig:static-potentials}
\end{figure}

Theoretical details of antistatic-antistatic potentials as well as their numerical computation with lattice QCD are extensively discussed in Refs.\ \cite{Wagner:2010ad,Bicudo:2015kna,Mueller:2023wzd,Bicudo:2024vxq}. In this work we use existing potentials from Refs.\ \cite{Bicudo:2015kna}, which were computed using $N_f = 2$ flavor ETMC gauge link ensembles \cite{ETM:2007xow,ETM:2009ztk} and extrapolated to physically light $u$ and $d$ quark masses. Relevant in the context of this work are the two $I = 0$ potentials $V_5(r)$ and $V_j(r)$ representing the interaction of two pseudoscalar and/or vector static light mesons. Suitable parameterizations of lattice QCD results for these potentials are
\begin{equation}
V_{X}(r) = -\frac{\alpha_{X}}{r} e^{-(r/d_X)^2} \quad , \quad X = 5,j
\end{equation}
with $\alpha_{5} = 0.34 \pm 0.03$, $d_{5} = 0.45^{+0.12}_{-0.10} \, \text{fm}$ (see Ref.\ \cite{Bicudo:2015kna}) and $\alpha_{j} = -0.10 \pm 0.07$ and \\ $d_{j} = (0.28 \pm 0.017) \text{fm}$ (see Ref.\ \cite{Bicudo:2016ooe}). The parameterizations are shown in Figure~\ref{fig:static-potentials}. For details we refer to Section~II of our recent publication \cite{Hoffmann:2024hbz}.


\section{\label{SEC_SG1}Coupled-Channel Schr\"odinger Equations}


\subsection{Identical Heavy Flavors: $\bar b \bar b u d$ with $I(J^P) = 0(1^-)$}

In Ref.\ \cite{Hoffmann:2024hbz} we have derived the coupled-channel Schr\"odinger equation relevant for the $\bar b \bar b u d$ system with $I(J^P) = 0(1^-)$. It is given by
\begin{equation}
	\left(\begin{pmatrix} 2 m_B & 0 \\ 0 & 2 m_{B^\ast} \end{pmatrix} - \frac{\nabla^2}{2 \mu_{bb}} + H_{\text{int}, S=0}\right) \vec{\varphi}_{L=1,S=0}(r) = E \vec{\varphi}_{L=1,S=0}(r)
	\label{eq:spin0-equation}
\end{equation}
with
\begin{equation}
	\nabla^2 = \frac{d^2}{d r^2} + \frac{2}{r} \frac{d}{d r} - \frac{L (L+1)}{r^2}\bigg|_{L = 1} = \frac{d^2}{d r^2} + \frac{2}{r} \frac{d}{d r} - \frac{2}{r^2}
	\label{EQN_nabla_sq}
\end{equation}
($L$ represents the orbital angular momentum of the heavy antiquarks) and
\begin{equation}
	H_{\text{int}, S=0} = \frac{1}{4} \begin{pmatrix} V_{5}(r) + 3 V_{j}(r) & \sqrt{3} (V_{5}(r) - V_{j}(r)) \\ \sqrt{3} (V_{5}(r) - V_{j}(r)) & 3 V_{5}(r) + V_{j}(r) \end{pmatrix} ,
	\label{eq:potential-matrix-2x2}
\end{equation}
where $\mu_{bb} = m_b/2$ is the reduced $b$ quark mass, $S = 0$ denotes the heavy spin and $r$ is the radial coordinate of the heavy antiquark separation. The 2 components of the wave function represent the following meson-meson combinations:
\begin{equation}
	\vec{\varphi}_{L=1,S=0} \equiv \bigg( BB \ \ , \ \ \frac{1}{\sqrt{3}} \vec{B}^\ast \vec{B}^\ast \bigg)^T = \bigg( BB \ \ , \ \ \frac{1}{\sqrt{3}} \Big(B_x^\ast B_x^\ast + B_y^\ast B_y^\ast + B_z^\ast B_z^\ast\Big)\bigg)^T .
	\label{EQN_phi}
\end{equation}


\subsection{Different Heavy Flavors: $\bar b \bar c u d$ with $I(J^P) = 0(0^+)$ and $I(J^P) = 0(1^+)$}

To derive the coupled-channel Schr\"odinger equations for the $\bar b \bar c u d$ systems one can closely follow Refs.\ \cite{Bicudo:2016ooe,Hoffmann:2024hbz} as sketched in the following.

$\bar{b} \bar{c} u d$ systems at large $\bar b \bar c$ separations $r$ are meson pairs, where one of the two mesons is a $B$ or $B^\ast$ meson and the other meson is a $D$ or $D^\ast$ meson. Consequently, the free Hamiltonian describing non-interacting meson pairs in the center of mass frame has a $16 \times 16$ matrix structure,
\begin{equation}
\label{EQN_H0} H_0 = M_{B} \tens{} \mathbb{1}_{4 \times 4} + \mathbb{1}_{4 \times 4} \tens{} M_D + \frac{\vec{p}^2}{2 \mu_{bc}} ,
\end{equation}
where $M_B = \mathrm{diag}(m_B , m_{B^\ast} , m_{B^\ast} , m_{B^\ast})$ and $M_D = \mathrm{diag}(m_D , m_{D^\ast} , m_{D^\ast} , m_{D^\ast})$ are diagonal matrices containing the meson masses and $\mu_{bc} = m_b m_c / (m_b + m_c)$ is the reduced mass of a $b$ and a $c$ quark. This Hamiltonian acts on a 16-component wave function for the relative coordinate of the heavy quarks $\vec{r}$, where the components can be interpreted as
\begin{eqnarray}
	\nonumber & & \hspace{-0.7cm} \vec{\Psi} \equiv \Big(
	B D     , B D_x^*     , B D_y^*     , B D_z^*     \ \ , \ \
	B_x^* D , B_x^* D_x^* , B_x^* D_y^* , B_x^* D_z^* \ \ , \ \
	B_y^* D , B_y^* D_x^* , B_y^* D_y^* , B_y^* D_z^* \ \ , \ \ \\
	\label{EQN003} & & \hspace{0.675cm} B_z^* D , B_z^* D_x^* , B_z^* D_y^* , B_z^* D_z^*\Big)^T .
\end{eqnarray}

To include interactions, one has to add the the potentials $V_5(r)$ and $V_j(r)$ discussed in Section~\ref{SEC_V5_Vj}. One can show that these potentials do not correspond to simple meson pairs, as represented by the components of $\vec{\Psi}$, but to linear combinations containing all four types of mesons, $B$, $B^\ast$, $D$ and $D^\ast$. These linear combinations can be expressed in terms of a $16 \times 16$ matrix $T$ using Fierz identities. The interacting part of the Hamiltonian is then
\begin{equation}
\label{EQN_Hint} H_{\text{int}} = T V_{\text{diag}} T^{-1} \quad , \quad V_{\text{diag}} = \text{diag}(\underbrace{V_5(r),\ldots,V_5(r)}_{4 \times} , \underbrace{V_j(r),\ldots,V_j(r)}_{12 \times}) .
\end{equation}
Combining Eq.\ (\ref{EQN_H0}) and Eq.\ (\ref{EQN_Hint}) leads to the Schr\"odinger equation
\begin{equation}
\label{EQN_SE_r_vec} H \vec{\Psi}(\vec{r}) = \Big(H_0 + H_\text{int}\Big) \vec{\Psi}(\vec{r}) = E \vec{\Psi}(\vec{r}) .
\end{equation}

Because $H_{\text{int}}$ only depends on the radial coordinate $r = |\vec{r}|$, but not on the direction of $\vec{r}$, the orbital angular momentum $L$ is conserved. Since the total angular momentum is a conserved quantity, the total spin $S$ is also conserved. Consequently, both $L$ and $S$ can be used as quantum numbers. The former allows to reduce the partial differential equation (\ref{EQN_SE_r_vec}) to an ordinary differential equation in $r$, while the latter allows to decompose the $16 \times 16$ Hamiltonian into smaller blocks.

We are particularly interested in $\bar{b} \bar{c} u d$ systems with quantum numbers $I(J^P) = 0(0^+)$ and $I(J^P) = 0(1^+)$, since recent full lattice QCD computations indicate the existence of shallow bound states for these systems \cite{Padmanath:2023rdu,Alexandrou:2023cqg,Radhakrishnan:2024ihu}. After working out the $I(J^P)$ quantum numbers for each block of the decomposed Hamiltonian using QCD symmetries and the Pauli principle (for details see e.g.\ Section~III.III in Ref.\ \cite{Hoffmann:2024hbz}), one can read off the relevant coupled channel Schr\"odinger equations.


\subsubsection*{Schr\"odinger Equation for $\bar{b} \bar{c} u d$ with $I(J^P)=0(0^+)$}

The coupled-channel Schr\"odinger equation for the $\bar{b} \bar{c} u d$ system with $I(J^P)=0(0^+)$ is
\begin{equation}
	\left(\begin{pmatrix} m_B + m_D & 0 \\ 0 &  m_{B^\ast} + m_{D^\ast} \end{pmatrix} - \frac{1}{2 \mu_{bc}} \frac{d^2}{d r^2} + H_{\text{int}, S=0}\right) \vec{\varphi}_{L=0,S=0}(r) = E \vec{\varphi}_{L=0,S=0}(r)
	\label{eq:bcud-0-plus-equation}
\end{equation}
with $H_{\text{int}, S=0}$ as defined in Eq.\ (\ref{eq:potential-matrix-2x2}). The 2 components of the wave function represent the following meson-meson combinations:
\begin{equation}
	\vec{\varphi}_{L=0,S=0} \equiv \bigg( BD \ \ , \ \ \frac{1}{\sqrt{3}} \vec{B}^\ast \vec{D}^\ast \bigg)^T = \bigg( BD \ \ , \ \ \frac{1}{\sqrt{3}} \Big(B_x^\ast D_x^\ast + B_y^\ast D_y^\ast + B_z^\ast D_z^\ast\Big)\bigg)^T .
\end{equation}


\subsubsection*{Schr\"odinger Equation for $\bar{b} \bar{c} u d$ with $I(J^P)=0(1^+)$}

The coupled-channel Schr\"odinger equation for the $\bar{b} \bar{c} u d$ system with $I(J^P)=0(1^+)$ is
\begin{equation}
	\left(\begin{pmatrix}
	m_{B^\ast} + m_D & 0 & 0 \\
	0 & m_B + m_{D^\ast} & 0 \\
	0 & 0 & m_{B^\ast} + m_{D^\ast}
	\end{pmatrix} - \frac{1}{2 \mu_{bc}} \frac{d^2}{d r^2} + H_{\text{int}, S=1}\right) \vec{\varphi}_{L=0,S=1,S_z}(r) = E \vec{\varphi}_{L=0,S=1,S_z}(r)
	\label{eq:bcud-0-plus-equation}
\end{equation}
with
\begin{equation}
	H_{\text{int}, S=1} = \frac{1}{4} \begin{pmatrix}
  V_5(r) + 3 V_j(r) & V_j(r) - V_5(r) & \sqrt{2} (V_5(r) - V_j(r)) \\
  V_j(r) - V_5(r) & V_5(r) + 3 V_j(r) & \sqrt{2} (V_j(r) - V_5(r)) \\
  \sqrt{2} (V_5(r) - V_j(r)) & \sqrt{2} (V_j(r) - V_5(r)) & 2 (V_5(r) + V_j(r))
	\end{pmatrix} .
\end{equation}
The 3 components of the wave function represent the following meson-meson combinations:
\begin{equation}
	\vec{\varphi}_{L=0,S=1,S_z} \equiv \Big(B^\ast_{S_z} D , B D^\ast_{S_z} , T_{1,S_z}(\vec{B}^\ast , \vec{D}^\ast)\Big)^T
\end{equation}
with $T_{1,S_z}$ denoting a spherical tensor coupling the three spin orientations of a $B^\ast$ and of a $D^\ast$ meson to a total spin $S=1$ with $z$ component $S_z$.


\section{\label{SEC_SG2}Scattering Formalism and the $\mbox{T}$ matrix}

Possibly existing bound states and resonances can be studied within the same formalism, by writing the wave function as a sum of an incident plane wave and an emergent spherical wave and by carrying out a partial wave expansion. Then, one can read off the $\mbox{T}$ matrix and determine its poles in the complex energy plane, which signal bound states or virtual bound states (for $\text{Re}(E_\text{pole}) < 2 m_B$ and $\text{Im}(E_\text{pole}) = 0$) or resonances (for $\text{Re}(E_\text{pole}) > 2 m_B$ and $\text{Im}(E_\text{pole}) < 0$). For details we refer to Section~IV of our recent publication~\cite{Hoffmann:2024hbz}.


\subsection{\label{SEC_T_for_bbud}The $\mbox{T}$ matrix for the $\bar b \bar b u d$ system with $I(J^P) = 0(1^-)$}

After the aforementioned partial wave expansion the $L = 1$ wave function (\ref{EQN_phi}) becomes
\begin{equation}
	\vec{\varphi}_{L=1,S=0}(r) = \begin{pmatrix}
	A_{B B} j_1(k_{B B} r) + \chi_{B B}(r) / r \\
	A_{B^\ast B^\ast} j_1(k_{B^\ast B^\ast} r) + \chi_{B^\ast B^\ast}(r) / r
	\end{pmatrix} ,
	\label{EQN_phi_decomposition}
\end{equation}
where $A_{B B}$ and $A_{B^\ast B^\ast}$ are the prefactors of the incident $B B$ and $B^\ast B^\ast$ waves, respectively, $j_1(k_{B B} r)$ and $j_1(k_{B^\ast B^\ast} r)$ denote spherical Bessel functions with scattering momenta $k_{B B} = \sqrt{2 \mu (E - 2m_B)}$ and $k_{B^\ast B^\ast} = \sqrt{2 \mu (E - 2m_{B^\ast})}$ representing the $L = 1$ contribution to these incident plane waves and $\chi_{B B}(r) / r$ and $\chi_{B^\ast B^\ast}(r) / r$ are the radial wave functions of the emergent $B B$ and $B^\ast B^\ast$ spherical waves. Inserting $\vec{\varphi}_{L=1,S=0}(r)$ from Eq.\ (\ref{EQN_phi_decomposition}) into the Schr\"odinger equation (\ref{eq:spin0-equation}) leads to
\begin{equation}
	\left(\begin{pmatrix} 2 m_B & 0 \\ 0 & 2 m_{B^\ast} \end{pmatrix} - \frac{1}{2 \mu_{bb}} \bigg(\frac{d^2}{d r^2} - \frac{2}{r^2}\bigg) + H_{\text{int}, S=0} - E\right)
	\begin{pmatrix} \chi_{B B}(r) \\ \chi_{B^\ast B^\ast}(r) \end{pmatrix} =
	-H_{\text{int}, S=0}
	\begin{pmatrix} A_{B B} r j_1(k_{B B} r) \\ A_{B^\ast B^\ast} r j_1(k_{B^\ast B^\ast} r) \end{pmatrix} .
	\label{EQN_SE_chi}
\end{equation}
As usual, the boundary conditions for the wave functions close to the origin are \\ $\chi_{\alpha}(r) \propto r^{L+1}|_{L=1} = r^2$.
For large $r$ the wave functions $\chi_\alpha(r)$ exclusively describe emergent spherical waves and, thus, are proportional to spherical Hankel functions,
\begin{eqnarray}
	\label{EQN_bc1} \chi_\alpha(r) \propto i r t_{B B;\alpha} h_1^{(1)}(k_\alpha r) & & \text{for } r \rightarrow \infty \text{ and } (A_{B B} , A_{B^\ast B^\ast}) = (1 , 0) \\
	\label{EQN_bc2} \chi_\alpha(r) \propto i r t_{B^* B^*;\alpha} h_1^{(1)}(k_\alpha r) & & \text{for } r \rightarrow \infty \text{ and } (A_{B B} , A_{B^\ast B^\ast}) = (0 , 1) ,
\end{eqnarray}
where $t_{\alpha;\beta}$ denote entries of the $\mbox{T}$ matrix. Thus, Eq.\ (\ref{EQN_bc1}) and Eq.\ (\ref{EQN_bc2}) allow to determine the $2 \times 2$ $\mbox{T}$ matrix,
\begin{equation}
	\mbox{T} = \begin{pmatrix}
	t_{B B ; B B} & t_{B B ; B^\ast B^\ast} \\
	t_{B^\ast B^\ast ; B B} & t_{B^\ast B^\ast ; B^\ast B^\ast}
	\end{pmatrix} .
	\label{eq:t-matrix-definition}
\end{equation}


\subsection{$\mbox{T}$ Matrices for the $\bar b \bar c u d$ Systems with $I(J^P) = 0(0^+)$ and $I(J^P) = 0(1^+)$}

One can proceed as sketched in Section~\ref{SEC_T_for_bbud}. Because $L = 0$ in both cases one has to replace $j_1$ by $j_0$. Moreover, scattering momenta have to be defined according to the meson types associated with each channel. At the end one arrives at a $2 \times 2$ $\mbox{T}$ matrix for $I(J^P) = 0(0^+)$ and at a $3 \times 3$ $\mbox{T}$ matrix for $I(J^P) = 0(1^+)$. Because of the page limit, we refrain from providing the corresponding equations.


\section{\label{SEC_results}Numerical Results}

The following numerical results were generated using quark masses $m_b = 4977 \, \text{MeV}$ and $m_c = 1628 \, \text{MeV}$ taken from a quark model \cite{Godfrey:1985xj}. For the meson mass splittings we use \\ $m_{B^\ast} - m_B = 45 \, \text{MeV}$ and $m_{D^\ast} - m_D = 138 \, \text{MeV}$ as quoted by the PDG \cite{ParticleDataGroup:2024cfk}. We solved the coupled-channel radial Schr\"odinger equations for the wave functions of the emergent wave $\chi_\alpha(r)$ for given complex energy $E$ using a standard fourth order Runge-Kutta integrator (e.g.\ Eq.\ (\ref{EQN_SE_chi}) in the case of the $\bar b \bar b u d$ system with $I(J^P) = 0(1^-)$). Then we read off the corresponding $\mbox{T}$ matrix elements from the behavior of the resulting $\chi_\alpha(r)$ at large $r$, using e.g.\ Eq.\ (\ref{EQN_bc1}) and Eq.\ (\ref{EQN_bc2}) for the $\bar b \bar b u d$ system with $I(J^P) = 0(1^-)$. Finally, we determine the poles of the $\mbox{T}$ matrix by numerically searching for roots of $1 / \text{det}(\mbox{T})$. For details we refer again to our recent publication \cite{Hoffmann:2024hbz}.


\subsection{$\bar{b} \bar{b} u d$ Tetraquark Resonance with $I(J^P) = 0(1^-)$}

Numerical results for the $\bar b \bar b u d$ system with $I(J^P) = 0(1^-)$ are extensively discussed in Ref.\ \cite{Hoffmann:2024hbz}. Our main findings are the following:
\begin{itemize}
\item[(1)] We found a pole of the $\mbox{T}$ matrix on the $(-,-)$-Riemann sheet
\footnote{For $n$ scattering channels there are $n$ scattering momenta $k_\alpha$ and $2^n$ Riemann sheets for the complex energy $E$. These sheets are labeled by the signs of the imaginary parts of the scattering momenta, e.g.\ by $(\text{sign}(k_{B B}) , \text{sign}(k_{B^\ast B^\ast}))$ for the $\bar{b} \bar{b} u d$ system with $I(J^P) = 0(1^-)$. There is a one-to-one correspondence between bound states and poles on the negative real axis of the physical Riemann sheet, which is characterized by having exclusively positive signs, e.g.\ the $(+,+)$ sheet in the case of 2-channel scattering. }
indicating a tetraquark resonance with mass $2 m_B + 94.0^{+1.3}_{-5.4} \, \text{MeV} = 2 m_{B^\ast} + 4.0^{+1.3}_{-5.4} \, \text{MeV}$, i.e.\ slightly above the $B^\ast B^\ast$ threshold, and decay width $\Gamma = 140^{+86}_{-66} \, \text{MeV}$.

\item[(2)] The coupled channel Schr\"odiger equation (\ref{EQN_SE_chi}), in particular the potential matrix (\ref{eq:potential-matrix-2x2}), led to a solid physical understanding, why there is a tetraquark resonance close to the $B^\ast B^\ast$ threshold, but not in the region of the $B B$ threshold, as naively expected from our previous work \cite{Bicudo:2017szl} using a simplified single-channel approach. The reason is that the attractive potential $V_5(r)$ dominates the $B^\ast B^\ast$ channel, but is strongly suppressed in the $B B$ channel, whereas the situation is reversed for the repulsive potential $V_j(r)$.

\item[(3)] This theoretical result is supported by our computation of branching ratios, where we found $\text{BR}_{B B} = 26^{+9}_{-4} \%$ and $\text{BR}_{B^\ast B^\ast} = 74^{+4}_{-9} \%$, implying that a decay of the tetraquark resonance is around three times more likely to a $B^\ast B^\ast$ pair than to a $B B$ pair.
\end{itemize}


\subsection{$\bar{b} \bar{c} u d$ virtual bound states with $I(J^P) = 0(0^+)$ and $I(J^P) = 0(1^+)$}


\subsubsection*{Virtual Bound States}

Using the same techniques as for the $\bar{b} \bar{b} u d$ tetraquark resonance with $I(J^P) = 0(1^-)$, we also searched for poles of the $\mbox{T}$ matrix in the complex energy plane for the two $\bar{b} \bar{c} u d$ systems. These pole searches were carried out on all four Riemann sheets for $I(J^P) = 0(0^+)$ and on all eight Riemann sheets for $I(J^P) = 0(1^+)$. For both systems we did neither find bound states nor resonances, but virtual bound states, indicated by poles on the negative real axis on the $(-,+)$-sheet and on the $(-,+,+)$-sheet, respectively. These poles are, however, rather far away from the lowest meson-meson thresholds, $\text{Re}(E) - (m_B + m_D) = -106^{+65}_{-148} \, \text{MeV}$ and $\text{Re}(E) - (m_{B^\ast} + m_D) = -100^{49}_{-212} \, \text{MeV}$. Thus, it is questionable, whether they have a sizable effect on physical observables like scattering rates or cross sections. We plan to investigate this in more detail in the near future.


\subsubsection*{Dependence on the Charm Quark Mass for $I(J^P) = 0(1^+)$}

\begin{figure}[htb]
    \centering
    \includegraphics[scale=0.6]{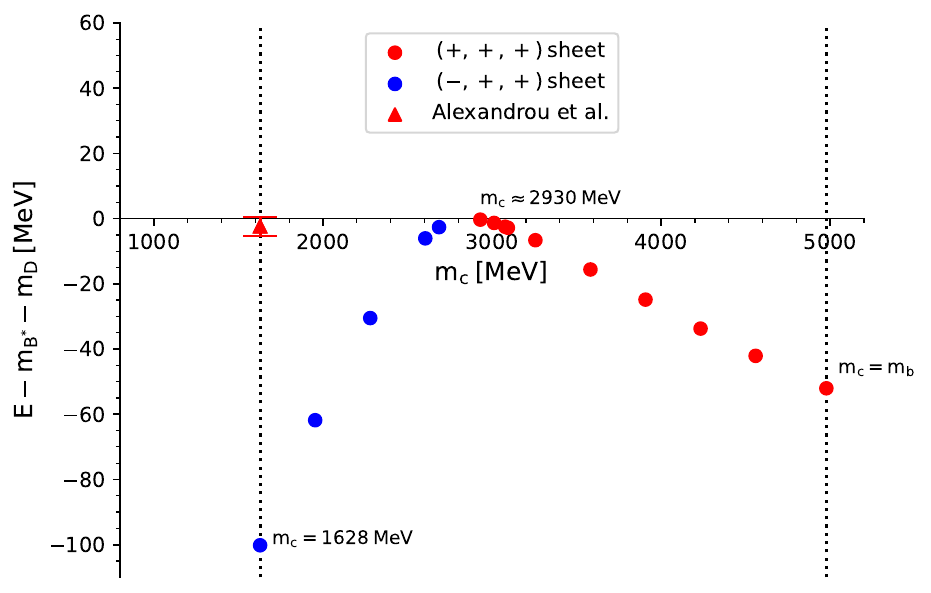}
    \caption{\label{fig:Tbc_E_vs_mc}The energy of the $\mbox{T}$ matrix pole as function of the charm quark mass $m_c$ for the $\bar{b} \bar{c} u d$ system with quantum numbers $I(J^P) = 0(1^+)$. The red triangular data point represents the full lattice QCD result from Ref.\ [10].}
\end{figure}

In Ref.\ \cite{Bicudo:2016ooe} we used the same potentials and formalism discussed in Section~\ref{SEC_V5_Vj} and Section~\ref{SEC_SG1} to predict a $\bar{b} \bar{b} u d$ bound state with quantum numbers $I(J^P)=0(1^+)$ and binding energy \\ $(m_B + m_{B^\ast}) - E = 59^{+30}_{-38} \, \text{MeV}$. This system, which has a $B B^\ast$ channel and a $B^\ast B^\ast$ channel is conceptually similar to the $\bar{b} \bar{c} u d$ system with the same quantum numbers. In particular, one can show that, when setting $m_c = m_b$, $m_D = m_B$ and $m_{D^\ast} = m_{B^\ast}$ in the coupled channel Schr\"odinger equation (\ref{eq:bcud-0-plus-equation}), one equation decouples and the remaining two equations are essentially identical to those solved in Ref.\ \cite{Bicudo:2016ooe}. We have studied this numerically by starting with Eq.\ (\ref{eq:bcud-0-plus-equation}) and physical quark masses $m_b$ and $m_c$ as provided at the beginning of Section~\ref{SEC_results} and then continously increasing $m_c$ from its physical value $1628 \, \text{MeV}$ to the value of the $b$ quark mass. At the same time we reduce the mass splitting between the $D$ and the $D^\ast$ meson according to
\begin{equation}
m_{D^\ast} - m_D = \frac{m_b}{m_c} \Big(m_{B^\ast} - m_B\Big) ,
\end{equation}
%
%
which is the leading order in Heavy Quark Effective Theory \cite{Neubert:1993mb}. The resulting pole energy as function of $m_c$ is shown in Figure~\ref{fig:Tbc_E_vs_mc}. One can see the expected transition from a virtual bound state corresponding to a pole on the $(-,+,+)$-Riemann sheet to a bound state corresponding to a pole on the physical $(+,+,+)$-Riemann sheet. The transition between the two sheets takes place at $m_c \approx 2930 \, \text{MeV}$, where the pole is located at $E = 0$. For $m_c = m_b$ we recover the binding energy $(m_B + m_{B^\ast}) - E = 59^{+30}_{-38} \, \text{MeV}$ predicted in Ref.\ \cite{Bicudo:2016ooe}, which is an excellent cross check and shows consistency of this work and Ref.\ \cite{Bicudo:2016ooe}.

\subsubsection*{Comparison to Full Lattice QCD Results}

Recent full lattice QCD studies of $\bar{b} \bar{c} u d$ systems with quantum numbers $I(J^P) = 0(0^+)$ and $I(J^P) = 0(1^+)$ have predicted shallow bound states rather close to the $B D$ threshold and the $B^\ast D$ threshold, respectively \cite{Padmanath:2023rdu,Alexandrou:2023cqg,Radhakrishnan:2024ihu} (the result from Ref.\ \cite{Alexandrou:2023cqg} for $I(J^P) = 0(1^+)$ is plotted in Figure~\ref{fig:Tbc_E_vs_mc}). It is interesting to note that the study from Ref.\ \cite{Alexandrou:2023cqg}, which uses a very advanced lattice QCD setup (large symmetric correlation matrices including both local and scattering interpolating operators, L\"uschers finite volume method to carry out a scattering analysis), cannot rule out the existence of shallow virtual bound states, even though genuine bound states are strongly favored. In any case there is a sizable quantitative difference of these full lattice QCD results and our $\bar{b} \bar{c} u d$ predictions from this work, which are based on lattice QCD potentials and the Born-Oppenheimer approximation. A possible reason for that could be that the attraction of the potential $V_5(r)$ was underestimated in Refs.\ \cite{Wagner:2010ad,Bicudo:2015kna}. To check this, we have recently started a recomputation of these potentials using a significantly improved up-to-date lattice QCD setup \cite{Mueller:2023wzd,Bicudo:2024vxq}.


\section*{Acknowledgements}

We acknowledge interesting and useful discussions with Pedro Bicudo.
J.H.\ acknowledges support by a ``Rolf and Edith Sandvoss Stipendium''.
M.W.\ acknowledges support by the Deutsche Forschungsgemeinschaft (DFG, German Research Foundation) -- project number 457742095.
M.W.\ acknowledges support by the Heisenberg Programme of the Deutsche Forschungsgemeinschaft (DFG, German Research Foundation) -- project number 399217702.
Calculations on the GOETHE-NHR and on the on the FUCHS-CSC high-performance computers of the Frankfurt University were conducted for this research. We thank HPC-Hessen, funded by the State Ministry of Higher Education, Research and the Arts, for programming advice.


\end{document}